\documentclass[letterpaper]{article}
\usepackage{aaai2027}
\usepackage[hyphens]{url}
\usepackage{graphicx}
\usepackage{natbib}
\usepackage{caption}
\usepackage{algorithm}
\usepackage{algorithmic}

\usepackage{newfloat}
\usepackage{listings}
\DeclareCaptionStyle{ruled}{labelfont=normalfont,labelsep=colon,strut=off}
\floatstyle{ruled}
\newfloat{listing}{tb}{lst}{}
\floatname{listing}{Listing}

\usepackage{booktabs}

\usepackage{amsmath}
\usepackage{amsthm}
\usepackage{amsfonts}

\usepackage{cleveref}

\theoremstyle{definition}
\newtheorem{definition}{Definition}

\usepackage{subcaption}

\nocopyright

\title{Stress-Relief Annealing: Polynomial-Time Simulation-Free Layout Optimization for Automated Warehouses}
\author {
    Xiangjie Luo\textsuperscript{\rm 1},
    Yulun Zhang\textsuperscript{\rm 2},
    Miyuki Koshimura\textsuperscript{\rm 1},
    Makoto Yokoo\textsuperscript{\rm 1},
    Jiaoyang Li\textsuperscript{\rm 2}
}
\affiliations {
    \textsuperscript{\rm 1}Graduate School of Information Science and Electrical Engineering, Kyushu University\\
    \textsuperscript{\rm 2}Robotics Institute, Carnegie Mellon University\\
    luo.xiangjie.239@s.kyushu-u.ac.jp, yulunzhang@cmu.edu, koshi@inf.kyushu-u.ac.jp, yokoo@kyudai.jp, jiaoyangli@cmu.edu
}

\begin{document}

\maketitle

\begin{abstract}
We study the problem of optimizing physical layouts for automated warehouses, where hundreds to thousands of robots are coordinated to transport packages. Previous works have shown that optimizing the warehouse layout (e.g., the physical location of the storage shelves) significantly improves throughput.
However, state-of-the-art layout optimization approaches are based on evolutionary optimization methods, which treat the entire warehouse as a black box and rely on random mutation to search for high-quality layouts.
While the optimization outcomes are promising, these methods require a massive number of simulations to evaluate candidate solutions, making them sample-inefficient.
In this paper, we present \textbf{S}tress-\textbf{R}elief \textbf{A}nnealing (SRA), a \emph{polynomial-time simulation-free} layout optimization algorithm. SRA turns the task demand into a per-vertex \emph{stress field} that predicts where traffic will concentrate in the warehouse; the field's peak provably caps the throughput.

Our experimental results show that (1) SRA improves both the throughput and the scalability of a human-designed warehouse, roughly doubling the number of robots it can sustain, (2) it matches or exceeds the throughput of the evolutionary baselines while taking only $19$ minutes on one CPU core, against their $25{,}000$ simulations and $25$ hours on a $64$-core machine, and (3) the gain generalizes across different Multi-Agent Path Finding algorithms, non-uniform task demands, and a warehouse with doubled dimensions.

\textit{Our code will be publicly available upon acceptance.}

\end{abstract}

\section{Introduction}
\label{sec:intro}

In an automated warehouse~\cite{wurman2008coordinating,KouAAAI20}, hundreds to thousands of robots are coordinated to continuously pick up goods from shelves and deliver them to workstations.
The key performance metric is throughput, computed as the number of tasks completed by all robots per timestep.
Where the robot traffic concentrates is set jointly by the Multi-Agent Path Finding (MAPF)~\cite{Stern2019benchmark} planner and the layout: the planner plans collision-free paths from the robots' start to goal locations, and the layout fixes which aisles they can use.
Under a fixed planner, the layout decides whether the traffic spreads out or funnels into bottlenecks---and with it, how many robots the system can sustain.

\begin{figure}[t!]
\centering
\includegraphics[width=0.9\columnwidth]{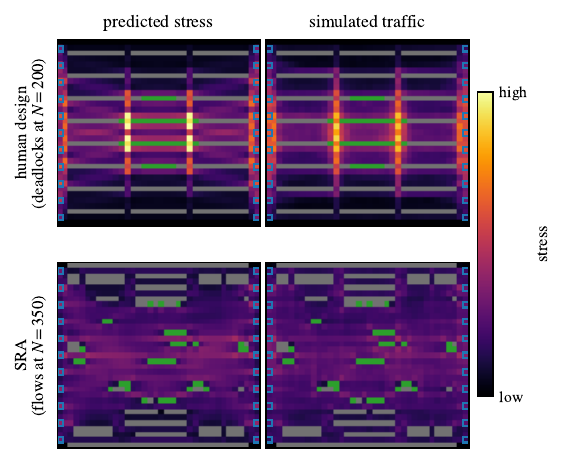}
\caption{The stress field is predicted from the shelf demand alone; the simulated traffic is one RHCR (PBS) run at $w=10$: $N=150$ on the original layout, $N=300$ on the SRA layout.}
\label{fig:teaser}
\end{figure}

Prior work has shown that optimizing the layout significantly improves throughput~\cite{zhangLayout23,ZhangNCA2023}.
However, state-of-the-art layout optimizers are evolutionary methods~\cite{bhatt2022dsage,fontaine2023cmamae}.
They treat the warehouse as a black box: candidate layouts are produced by random mutation, and each candidate is evaluated by a computationally expensive multi-robot simulation that returns the throughput.
These methods are therefore sample-inefficient: \citet{zhangLayout23} reports optimizing a $33\times36$ layout with $200$ robots on a $64$-core machine in $24$ hours. At real-world scales---up to $4{,}000$ robots in a single warehouse~\cite{Brown2023amazonrobot} and maps as large as $179\times69$~\cite{Yu2023}---simulation-driven search only becomes less affordable.
Moreover, the \emph{shelf demand}---how often each shelf is requested---is skewed in practice: a small set of \emph{high-frequency} shelves absorbs most of the requests, while the \emph{low-frequency} majority is rarely visited.
While the evolutionary methods can in principle be run under such non-uniform demand, their search is not informed by it: random mutation does not know which shelves are requested frequently, so placing the high-frequency shelves well is left to chance.

We close this gap with a cheap congestion estimate that is explicitly informed by the shelf demand.
From this demand alone, we compute a \emph{stress field}: an estimate of how the robot traffic will spread over the warehouse, obtained by repeatedly running single-agent planning between shelves and workstations---no multi-robot simulation involved.
The field serves as a \emph{surrogate} for simulation: it predicts a candidate layout's congestion without running simulations.
\Cref{fig:teaser} previews the idea: the field, computed before any robot moves, marks the same corridors that later become congested in simulation.

Building on this surrogate, we present \textbf{S}tress-\textbf{R}elief \textbf{A}nnealing (SRA), which repeats a simple cycle: compute the stress field of the current layout, then relocate shelves away from the most stressed regions, accepting or rejecting each change by simulated annealing~\cite{kirkpatrick1983optimization} (Sec.~\ref{sec:sra}).
The procedure calls no simulation and provably runs in polynomial time.

We validate SRA in simulation with the state-of-the-art lifelong MAPF algorithms.
Our experimental results show that (1) SRA improves both the throughput and the scalability of a human-designed warehouse, roughly doubling the number of robots it can sustain, (2) SRA matches or exceeds the throughput of the evolutionary baselines---which consume $25{,}000$ simulations and $25$ hours per run on average using 64 CPU cores in parallel---in $19$ minutes on one CPU core, and (3) the gain generalizes across different MAPF planners---the search-based RHCR~\cite{li2021lifelong} and the rule-based PIBT~\cite{okumura2022priority}---across non-uniform demand, and to a warehouse with doubled dimensions.
The gain generalizes because the field is faithful: the predicted field closely matches the traffic realized in simulation (Sec.~\ref{sec:exp-faithful}).

\section{Problem Formulation}
\label{sec:problem}

Following \citet{zhangLayout23}, we first define warehouse layout and then the problem of layout optimization.

\begin{figure}[t]
\centering
\includegraphics[width=0.8\columnwidth]{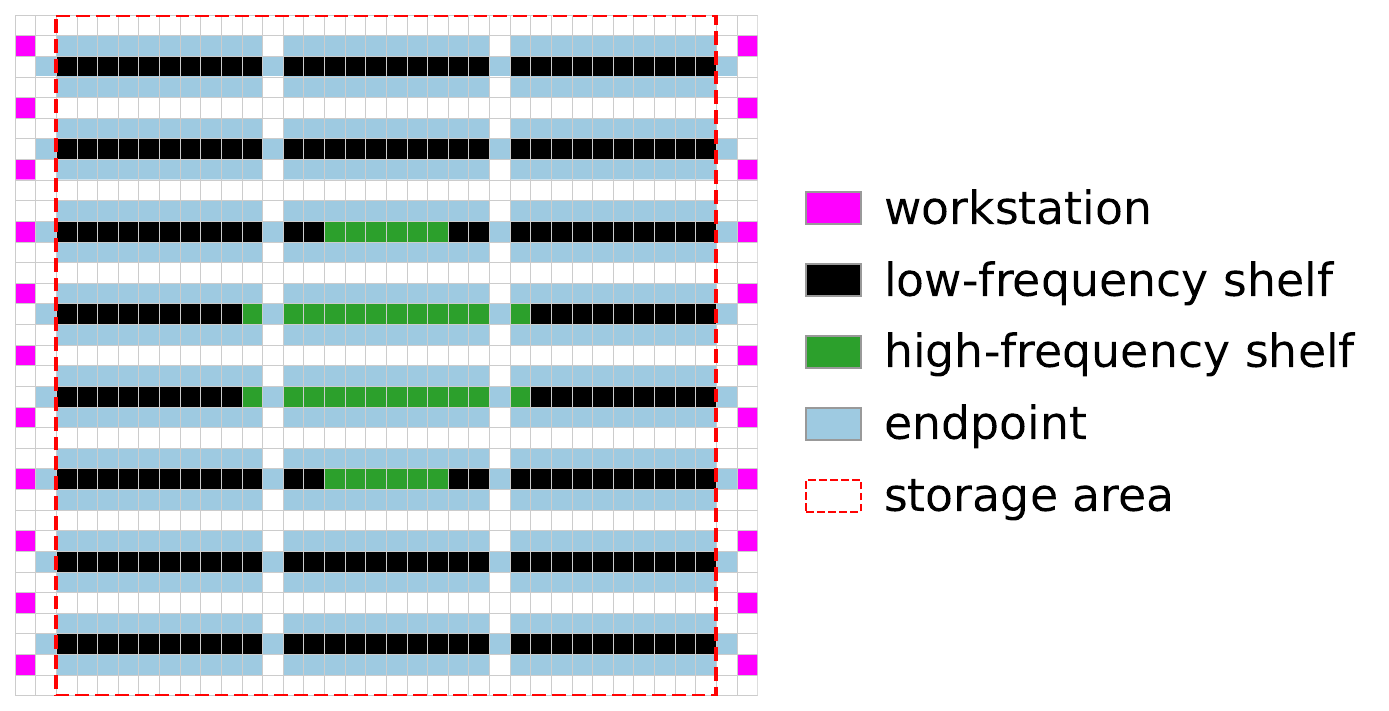}
\caption{The human-designed original layout.
Shelves inside the red dashed box (storage area) are the decision variables;
endpoints are the traversable vertices adjacent to each shelf.}
\label{fig:layout}
\end{figure}

\begin{definition}[Warehouse layout]
\label{def:layout}
A warehouse layout is a four-neighbor grid graph $G$, with $V$ being the set of all vertices. Every vertex is a shelf, a workstation, an endpoint, or an empty space. Robots can traverse non-shelf vertices.
\Cref{fig:layout} shows an example.
The grid splits into a \emph{storage area}, which holds all shelves, and a fixed \emph{non-storage margin}, which holds the workstations and is not optimized.
The shelf set is $S$, and the workstation set is $W$ with $|W|=M$.
A traversable vertex adjacent to a shelf $s \in S$ is an \emph{endpoint} of $s$, where a robot parks to interact with the shelf.
We write $\partial s$ for the endpoint set of $s \in S$.
Endpoints are induced by shelf placement---every traversable face of a shelf is one of its endpoints.
\end{definition}

Note that prior works~\cite{zhangLayout23,ZhangNCA2023} optimize the locations of both endpoints and shelves; our decision variables are the shelf positions alone.

\begin{definition}[Valid layout]
\label{def:valid}
A layout is valid iff (1) the traversable vertices form a single connected component, and (2) every shelf has at least one adjacent endpoint.
\end{definition}

Condition (1) makes $G$ connected, and condition (2) keeps every shelf accessible.

\paragraph{Robot Tasks.}
The task structure simulates an automated fulfillment warehouse~\cite{honig2019warehouse,zhangLayout23,li2021lifelong}: the goals of each robot alternate between endpoints and workstations---a robot picks up a package at a shelf's endpoint, delivers at a workstation, and then heads to the next endpoint.

\begin{definition}[Non-uniform shelf demand]
\label{def:demand}
Each shelf $s$ carries a fixed demand weight $w_s$, where $w_s=w$ for high-frequency shelves and $w_s=1$ for low-frequency ones; we call $w$ the \emph{demand skew}.
When a goal is sampled from the endpoints, it is drawn by weight: shelf $s$ splits $w_s$ equally over its endpoints, an endpoint shared by several shelves sums the shares, and an endpoint is drawn with probability proportional to its weight. The visit rate of a shelf is $w_s$ no matter how many faces it exposes.
When a goal is sampled from the workstations, workstation $j$ is drawn with probability $\omega_j$, where $\sum_j \omega_j = 1$.
$\{w_s\}$ and $\{\omega_j\}$ together form the \emph{demand structure}.
\end{definition}

\begin{definition}[Throughput]
\label{def:throughput}
A robot finishes a task when it reaches its currently assigned goal.
Throughput $\lambda$ is the average number of finished tasks per timestep.
We write $\bar{\lambda}(G,N)$ for the throughput of layout $G$ operated with $N$ robots under a fixed planner, and call $\lambda^*(G)=\max_N \bar{\lambda}(G,N)$ the \emph{peak throughput} of $G$ under that planner.
\end{definition}

\paragraph{Congestion collapse and capacity.}
Throughput grows with $N$, but a layout that funnels flow into a few aisles has a critical point: past it, congestion feeds on itself, the system deadlocks, and throughput drops to zero~\cite{li2021lifelong,zhangLayout23}.
For example, the layout of \Cref{fig:teaser} (top) sustains about 5 tasks per timestep at $N=150$; at $N=200$ eight of ten runs collapse (Sec.~\ref{sec:exp}).
The peak throughput $\lambda^*$ is therefore attained just before this collapse.
Measuring throughput requires a computationally expensive multi-robot simulation (Sec.~\ref{sec:lifelong}), and the prior layout optimization methods need a massive number of such evaluations~\cite{zhangLayout23,ZhangNCA2023}.

\begin{definition}[Layout optimization]
\label{def:problem}
Given the storage area, the shelf counts ($n_H$ high-frequency and $n_C$ low-frequency, both fixed), the demand structure $\{w_s\},\{\omega_j\}$, and the robot count $N$, layout optimization asks for a valid layout $G$ that maximizes the throughput $\bar{\lambda}(G,N)$.
\end{definition}

\section{Preliminaries}
\label{sec:prelim}
\subsection{Lifelong MAPF}
\label{sec:lifelong}

Multi-Agent Path Finding (MAPF)~\cite{Stern2019benchmark} plans collision-free paths for a team of agents from their corresponding start to goal locations.
Lifelong MAPF~\cite{ma2017lifelong,li2021lifelong} is a variant of MAPF that continuously assigns new goals to agents.
In this work, we adopt the classical lifelong MAPF definition for the operation of the robots, where (1) all robots move in discrete timesteps on a warehouse layout, (2) at each timestep, each robot can either stay at its current vertex or move to an adjacent vertex, (3) robots cannot be in the same vertex or swap vertices at the same timestep, and (4) the objective is to maximize throughput.

Lifelong MAPF algorithms broadly fall into two categories.
Search-based methods~\cite{WanICARCV18,li2021lifelong} decompose the problem into a sequence of MAPF problems and solve each using a MAPF planner. These methods produce high-quality solutions but lack scalability.
Rule-based methods~\cite{WangB11,WangICAPS08,okumura2022priority,Yu2023} decide the robots' moves one step at a time with a pre-defined rule, trading solution-quality guarantees for speed and scale.
We conduct experiments with RHCR~\cite{li2021lifelong} and PIBT~\cite{okumura2022priority}, the state-of-the-art representatives of the two categories.

\subsubsection{Congestion Approximation in Lifelong MAPF}
\label{sec:congestionapprox}
Several lines of work inside lifelong MAPF approximate where congestion will arise and use the result to steer the robots.
\citet{han2022space} penalizes the vertices occupied by the already-planned paths, so that later single-agent paths spread over the map.
\citet{chen2024traffic} repeatedly runs single-agent planning with congestion-aware edge costs and turns the resulting flows into guidance.
\citet{Yu2023} predicts congestion for large robot fleets with data-driven methods.
\citet{ewing2022betweenness} computes the betweenness centrality of the map---the construction closest to our stress field---and relates it to the empirical hardness of the map's MAPF instances.
The stress field of Sec.~\ref{sec:method} likewise aggregates single-agent shortest paths into a per-vertex congestion surrogate.
It differs in what the surrogate is for: these works steer the robots or predict solver performance, both on a fixed layout, whereas we use it to optimize the layout itself and tie its peak to a throughput bound.

\subsection{Warehouse Layout Optimization}
\label{sec:layoutopt}

Operations research evaluates warehouse layouts analytically: queuing-network models estimate the throughput of regular, template-like layouts~\cite{lamballais2017estimating,wu2020research}, and integer programming places workstations to minimize average travel distance~\cite{yang2021non}.
These models need no simulation, but their travel times are free-flow: robots do not interfere with each other, so the congestion collapse of Sec.~\ref{sec:problem} is invisible to them.
They also keep the shelves fixed: the layout follows a fixed template, and only a few design choices---the length-to-width ratio of the storage area, the workstation locations---are tuned.

The state-of-the-art layout optimization methods---DSAGE~\cite{zhangLayout23} and NCA~\cite{ZhangNCA2023}---are based on evolutionary optimization~\cite{mouret2015illuminating,Zhang2021DeepSA,fontaine2023cmamae}: mutation proposes candidate layouts and a lifelong MAPF simulation scores each one, which makes them impractical at real-world scales. SRA requires no lifelong MAPF simulation during optimization.

The analytical models and the evolutionary methods have complementary weaknesses: the former need no simulation but are blind to traffic congestion, while the latter are sensitive to congestion but pay for every candidate with a simulation.
The stress field of Sec.~\ref{sec:method} combines their strengths: an analytical congestion surrogate that requires no simulation.

\section{Method}
\label{sec:method}

\subsection{Stress Field and the Bottleneck Law}
\label{sec:surrogate}
We turn the demand structure of Definition~\ref{def:demand} into a per-vertex stress field.
Each shelf--workstation pair $(s,j)$ carries a flow of volume $w_s\,\omega_j$ from the endpoints of $s$ to workstation $j$; we spread this flow evenly over their shortest paths and superpose all pairs.

Let $F: V \rightarrow \mathbb{R}$ be the flow function. Then the flow at vertex $v \in V$ is:
\begin{equation} \label{eq:F}
F(v) = \sum_{s\in S}\sum_{j\in W} w_s\,\omega_j\, b_{s,j}(v).
\end{equation}
$b_{s,j}(v)$ is the average, over the endpoints of $s$, of the fraction of shortest paths from each endpoint to workstation $j$ that pass through $v$:
\begin{align}
b_{s,j}(v) &= \frac{1}{|\partial s|}\sum_{e\in\partial s}\frac{\sigma_{ej}(v)}{\sigma_{ej}}, \label{eq:b} \\
\sigma_{ej}(v) &=
\begin{cases}
\sigma_{ev}\,\sigma_{vj}, & d(e,v)+d(v,j)=d(e,j),\\
0, & \text{otherwise.} \label{eq:sigma}
\end{cases}
\end{align}
In \Cref{eq:b}, $\partial s$ is the set of adjacent endpoints of shelf $s$ (Definition~\ref{def:layout}); the prefactor $1/|\partial s|$ keeps the total contribution of $s$ at $w_s$ regardless of how many endpoints it exposes.
$\sigma_{ej}$ is the number of shortest paths between $e$ and $j$, $\sigma_{ej}(v)$ the number of such paths that pass $v$, and $d$ the shortest-path distance; the ratio $\sigma_{ej}(v)/\sigma_{ej}$ is the probability that a robot on a uniformly random shortest path from $e$ to $j$ passes through $v$.
\Cref{eq:sigma} is the standard shortest-path counting identity: $v$ lies on some shortest $e\to j$ path iff $d(e,v)+d(v,j)=d(e,j)$.

Therefore, $F$ in \Cref{eq:F} is a variant of vertex betweenness centrality~\cite{freeman1977betweenness} in which the sum runs only over endpoint--workstation pairs, each weighted by its demand $w_s\,\omega_j$; one BFS per workstation, followed by a Brandes-style dependency accumulation~\cite{brandes2001faster}, can compute it at total cost $O(M\,|V|)$.
Normalizing $F$ to unit demand gives the \emph{stress field} $\ell$:
\begin{equation}
\ell(v) = \frac{F(v)}{W_{\text{total}}},
\qquad
W_{\text{total}} = \sum_{s\in S} w_s .
\end{equation}
$\ell(v)$ is the probability that a random task's route passes through $v$; a shortest path crosses $v$ at most once, so this probability also equals the expected number of times that $v$ is traversed per task.
The denominator $W_{\text{total}}$ is determined by the demand structure alone; a layout-independent denominator keeps the optimizer from diluting the congestion reading by lengthening paths globally.

The construction above assumes that all traffic follows shortest routes---the all-or-nothing assignment of traffic analysis~\cite{wardrop1952road,sheffi1985urban}---ignoring that robots may detour around congestion.
Appendix~A.1 tests where this construction's guidance holds and finds it faithful over the robot counts studied here.

Congestion concentrates at the most loaded vertex.
We define the \emph{bottleneck load}:
\begin{equation}
\ell^* = \max_v \ell(v).
\end{equation}
At throughput $\lambda$, the traversal demand at $v$ is $\lambda\,\ell(v)$.
A \emph{traversal} of a vertex is one robot entering it.
A vertex holds one robot at a time, so the traversals it sustains per timestep are bounded; we write $c$ for the highest traversal rate a vertex sustains.
We model $c$ as a constant of the MAPF algorithm, independent of the layout, the robot count, and the demand structure, which enter only through $\ell(v)$.
We calibrate it once, as the highest per-vertex traversal rate observed before collapse, over RHCR runs spanning layouts and robot counts (Sec.~\ref{sec:exp-faithful} gives the value used here).
Sustaining throughput $\lambda$ therefore requires $\lambda\,\ell(v)\le c$ at every vertex, the tightest at the most loaded one, which gives the \emph{bottleneck law}
\begin{equation}
\lambda\,\ell^* \le c
\quad\Longrightarrow\quad
\lambda \le \frac{c}{\ell^*}.
\end{equation}
This is the classical bottleneck bound of queuing analysis~\cite{lazowska1984quantitative}, applied to the vertices of the grid.
Lowering $\ell^*$ raises the throughput ceiling, giving the basis for optimizing $\ell^*$.
Only this analysis needs the value of $c$: lowering $\ell^*$ raises the ceiling whatever $c$ is, so the optimizer of Sec.~\ref{sec:sra} never uses it.
In our experiments, layouts with a lower $\ell^*$ attain a higher peak throughput (Sec.~\ref{sec:exp}), the qualitative prediction of the bound.

SRA optimizes $\ell^*$ jointly with the \emph{weighted trip length}
\begin{equation}
\label{eq:Lw}
L_w = \frac{1}{W_{\text{total}}}\sum_{s\in S}\sum_{j\in W} w_s\,\omega_j\,\frac{1}{|\partial s|}\sum_{e\in\partial s} d(e,j).
\end{equation}
Given a random task drawn as in Definition~\ref{def:demand}, $L_w$ is the expected shortest-path distance between the endpoint and the workstation of the task.
The optimization lowers $\ell^*$ while keeping $L_w$ small.
All these quantities depend on the layout.
When comparing layouts, we write $\ell^*(G)$ and $L_w(G)$.

\subsection{Stress-Relief Annealing}
\label{sec:sra}

A congestion peak in a layout is a stress concentration, and the standard process for removing stress concentrations is annealing~\cite{kirkpatrick1983optimization}, which names the method SRA (Stress-Relief Annealing).
Algorithm~\ref{alg:sra} lists the full procedure; the rest of this section walks through it.
The two objectives combine into a single energy
\begin{equation} \label{eq:Energy}
E(G) = \alpha\,\ell^*(G) + L_w(G),
\end{equation}
where the \emph{congestion term} $\alpha\,\ell^*$ penalizes a high bottleneck load, and the \emph{trip term} $L_w$ penalizes long trips.
$\alpha$ sets their relative strength.
A fixed $\alpha$ drifts across maps and demand structures, because $\ell^*$ and $L_w$ differ in magnitude across $w$ and scale.
To compute $\alpha$, we first set $K$, the desired ratio of the initial congestion term $\alpha\ell^*_0$ to the initial trip term $L_{w,0}$ on the original layout, and derive $\alpha$ from it (lines 1--2 of Algorithm~\ref{alg:sra}):
\begin{equation}
\alpha = K\,\frac{L_{w,0}}{\ell^*_0},
\end{equation}
We set $K=1$: the two terms start equal.
$\alpha$ is computed once at the start and frozen throughout; $K$ transfers across maps and demand structures without per-setting tuning.

SRA iteratively selects a shelf and relocates it to a different position to optimize $E$.
At each step, SRA relocates one shelf in two stages: (1) select a source shelf (line 6), and (2) select a target vertex (line 7).
While selecting the source shelf, we first score every shelf $s$ by its \emph{felt stress}:

\begin{equation}
\mathrm{felt}(s) = \alpha\,\max_{e\in\partial s}\rho(e) + w_s\, \frac{1}{|\partial s|}\sum_{e\in\partial s} D(e).
\end{equation}
Here $\rho(e) = N\,F(e)/\sum_{u\in V} F(u)$ denotes the expected occupancy of $e$, i.e., the expected number of robots at $e$ when the $N$ robots are distributed over the map in proportion to the flow $F$ (\Cref{eq:F}).
$D(e)$ is the mean distance from $e$ to the workstations.
Both terms aggregate over the endpoints of $s$: the maximum for $\rho$ and the average for $D$.

The source is then sampled from a softmax over the z-scored felt values (\texttt{getSource}, Appendix~B): high-stress shelves are chosen more often, and the randomness breaks cyclic stalls.

We then select the target vertex. Since endpoints are induced by adjacency (Definition~\ref{def:layout}), a shelf may be relocated to any non-shelf vertex in the storage area, including vertices that currently serve as endpoints of other shelves.
We try the non-shelf vertices in the storage area in ascending order of placement cost:
\begin{equation}
\mathrm{cost}(v) = \alpha\,\rho(v) + w_s\, D(v),
\end{equation}
where $w_s$ is the demand weight of the relocated shelf.

Each candidate target vertex must pass the \emph{veto}: the layout must remain valid after relocating the source shelf to the candidate target (Definition~\ref{def:valid}).
The first vertex that passes is the proposal (\texttt{getTarget}, Appendix~B).
Acceptance of the relocation proposed at step $k$ follows the Metropolis rule~\cite{metropolis1953equation} (lines 8--12):
\begin{equation}
\label{eq:accept}
P_{\text{accept}} = \min\!\big(1,\ e^{-\Delta E / T_k}\big),\qquad
\Delta E = E(G') - E(G),
\end{equation}
where $G'$ is the layout after the relocation proposed at step $k$ and $G$ the current layout, under geometric cooling $T_{k+1}=\gamma T_k$ (line 16) with $\gamma=(0.01)^{1/\text{steps}}$, from $T_0$ to $0.01\,T_0$.
\texttt{initializeTemperature} (line 3) calibrates $T_0=\overline{|\Delta E|}$ over trial relocations that are scored but never applied (Appendix~B).
Line 4 initializes the energy and the best layout found so far.
During annealing, SRA keeps track of the lowest-energy layout encountered and returns it (lines 13--15 and 18).

\begin{algorithm}[t]
\caption{Stress-Relief Annealing (SRA)}
\label{alg:sra}
\begin{algorithmic}[1]
\REQUIRE original layout $G$, demand structure $\{w_s\},\{\omega_j\}$, robot count $N$, step count \textit{steps}, ratio $K$
\STATE $\ell^*_0,\ L_{w,0} \gets \texttt{computeStressField}(G)$
\STATE $\alpha \gets K\,L_{w,0}/\ell^*_0$
\STATE $T \gets \texttt{initializeTemperature}(G)$ \COMMENT{$T$: annealing temperature}
\STATE $E \gets \alpha\,\ell^*_0 + L_{w,0}$;\quad $G^\star \gets G$;\quad $E^\star \gets E$ \COMMENT{$G^\star,E^\star$: best so far}
\FOR{$k = 1$ \TO \textit{steps}}
  \STATE $s \gets \texttt{getSource}(G)$
  \STATE $v \gets \texttt{getTarget}(G, s)$
  \STATE $G' \gets G$ with $s$ relocated to $v$
  \STATE $\ell^*(G'),\ L_w(G') \gets \texttt{computeStressField}(G')$;\quad $E' \gets \alpha\,\ell^*(G') + L_w(G')$
  \IF{$\mathrm{rand}() < P_{\text{accept}}$}
    \STATE $G \gets G'$;\quad $E \gets E'$
  \ENDIF
  \IF{$E < E^\star$}
    \STATE $G^\star \gets G$;\quad $E^\star \gets E$
  \ENDIF
  \STATE $T \gets \gamma\,T$
\ENDFOR
\RETURN $G^\star$
\end{algorithmic}
\end{algorithm}

\paragraph{Runtime Complexity.}
Our implementation recomputes the stress field per endpoint, at $O(|S|\,M\,|V|)$ per step; the Brandes-style accumulation of Sec.~\ref{sec:surrogate} would lower this to $O(M\,|V|)$, which we have not implemented.
The felt and cost scores are read on the same pass, and each veto check is a connectivity test in $O(|V|)$.
In the worst case every empty vertex of the storage area is checked before one passes, and a run of \textit{steps} iterations costs $O(\textit{steps}\cdot(|S|\,M\,|V|+|V|^2))$, still polynomial.
The bound involves no lifelong MAPF simulation.

\section{Experiments}
\label{sec:exp}
\subsection{Setup}
\label{sec:setup}

Following prior works~\cite{ZhangNCA2023,zhangLayout23}, experiments use the warehouse of \Cref{fig:layout}: a $33\times36$ layout with $M=22$ workstations on the left and right margins and $240$ shelves ($36$ high-frequency, $204$ low-frequency) in the storage area.
The original layout is a commonly used human-designed layout~\cite{li2021lifelong,LiuAAMAS19}, with bottleneck load $\ell^*_0=0.139$.
The scaling experiment doubles both dimensions of this layout, keeping its aisle structure and its shelf and workstation densities: $66\times69$, $960$ shelves ($144$ high-frequency), $M=44$, with $\ell^*_0=0.084$ (construction detailed in Appendix~C).
We compare SRA with DSAGE~\cite{zhangLayout23} and NCA~\cite{ZhangNCA2023} (parameters detailed in Appendix~D) across demand skews $w\in\{1,5,10\}$ and across lifelong MAPF algorithms: RHCR~\cite{li2021lifelong} with PBS~\cite{ma2019searching} as its solver, and PIBT~\cite{okumura2022priority}.

\paragraph{SRA parameters.}
SRA optimizes for $N=300$ robots and runs $3500$ steps with $K=1$ and the calibrated $T_0$, on a single CPU core, in $18.8 \pm 0.4$ minutes per run.
We anneal five seeds at $w=10$ and three at each other skew.
\paragraph{Evaluation.}
Each optimized layout is evaluated by simulation over $1000$ timesteps.
The planners are RHCR and PIBT.
RHCR replans every $5$ timesteps with a planning window of $10$, using PBS as its solver throughout.
Unless stated otherwise, evaluations run at $N=300$ with $10$ simulation seeds.
The robot-count sweeps span $N=100$ to $400$ on the original warehouse and $N=300$ to $1500$ on the $66\times69$ warehouse.
Reported SRA results pool the annealing seeds unless stated otherwise.

\subsection{Main Results}
\label{sec:exp-main}

\begin{table}[t]
\centering
\caption{Throughput at $N=300$, and search cost per run.
Original is the human-designed original layout.
Each method's best layout is evaluated over $10$ simulation seeds: mean $\pm$ one standard deviation.
In the PIBT block, DSAGE and NCA use the best layouts of their searches re-run with PIBT evaluations; SRA uses the same layouts as in the RHCR block.
Time is mean $\pm$ $95\%$ CI over search runs.}
\label{tab:external}
\resizebox{\linewidth}{!}{
\begin{tabular}{lccccc}
\toprule
 & \multicolumn{3}{c}{$\lambda$ @ $N{=}300$} & \multicolumn{2}{c}{search cost} \\
\cmidrule(lr){2-4}\cmidrule(lr){5-6}
Method & $w{=}1$ & $w{=}5$ & $w{=}10$ & sims & time \\
\midrule
\multicolumn{6}{l}{\emph{RHCR (PBS)}} \\
Original & $0.73 \pm 0.32$ & $0.24 \pm 0.10$ & $0.13 \pm 0.06$ & -- & -- \\
DSAGE & $8.23 \pm 0.90$ & $7.64 \pm 1.87$ & $8.26 \pm 1.16$ & 25000 & $19.2 \pm 1.0$\,h \\
NCA   & $\mathbf{9.24 \pm 0.06}$ & $8.76 \pm 1.50$ & $8.53 \pm 1.98$ & 25000 & $29.6 \pm 1.1$\,h \\
SRA   & $\mathbf{9.25 \pm 0.05}$ & $\mathbf{9.58 \pm 0.06}$ & $\mathbf{9.81 \pm 0.04}$ & \textbf{0} & $\mathbf{18.8 \pm 0.4}$\,min \\
\midrule
\multicolumn{6}{l}{\emph{PIBT}} \\
Original & $4.77 \pm 0.06$ & $4.76 \pm 0.06$ & $4.45 \pm 0.14$ & -- & -- \\
DSAGE & $5.96 \pm 0.09$ & $6.59 \pm 0.07$ & $6.52 \pm 0.09$ & 25000 & $2.5 \pm 0.3$\,h \\
NCA   & $\mathbf{7.11 \pm 0.05}$ & $7.06 \pm 0.03$ & $7.14 \pm 0.04$ & 25000 & $2.6 \pm 0.1$\,h \\
SRA   & $\mathbf{7.16 \pm 0.06}$ & $\mathbf{7.18 \pm 0.05}$ & $\mathbf{7.77 \pm 0.06}$ & \textbf{0} & $\mathbf{18.8 \pm 0.4}$\,min \\
\bottomrule
\end{tabular}
}
\end{table}

\begin{figure}[t]
\centering
\includegraphics[width=0.55\columnwidth]{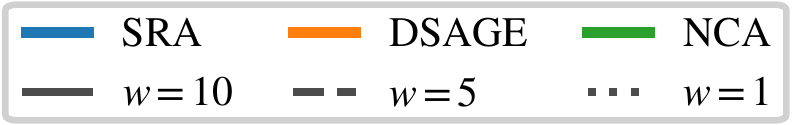}\\
 \begin{subfigure}{0.24\textwidth}
    \includegraphics[width=1\textwidth]{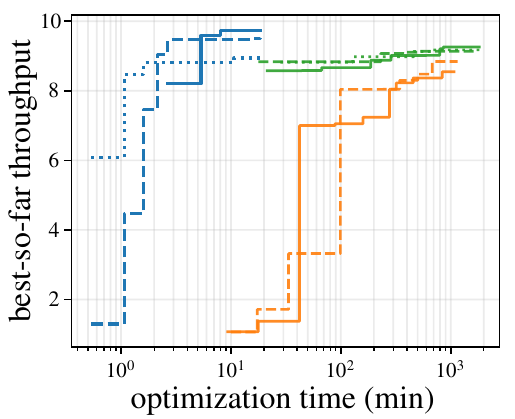}
 \caption{RHCR}
\end{subfigure}%
\begin{subfigure}{0.24\textwidth}
    \includegraphics[width=1\textwidth]{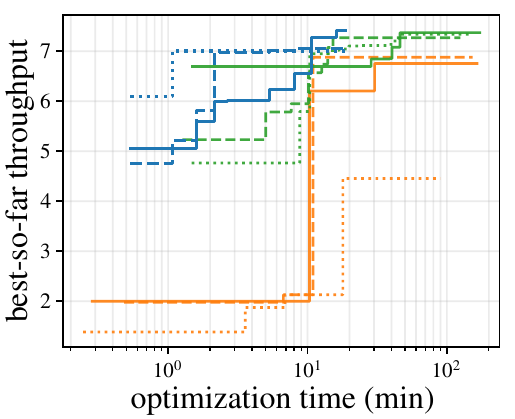}
 \caption{PIBT}
 \end{subfigure}

\caption{Best-so-far throughput against optimization wall-clock time (log scale) for RHCR and PIBT at $w \in \{1,5,10\}$.
SRA curves are means over annealing seeds, $10$ RHCR (PBS) and $5$ PIBT simulation seeds at $N=300$ per snapshot, on one CPU core. DSAGE and NCA curves are the best of their search runs per skew (run set and time normalization in Appendix~D).}
\label{fig:convergence}
\end{figure}

\paragraph{Comparison with evolutionary search.}
Table~\ref{tab:external} compares SRA with DSAGE and NCA on best-layout throughput.
SRA matches the best baseline at $w=1$ and outperforms both baselines at $w=5$ and $w=10$; the advantage grows with the demand skew.
Every SRA run ends above $\lambda=9.1$, while DSAGE in every skew and NCA in $w=5$ and $w=10$ have runs that end at $\lambda=2.9$--$5.9$.
\begin{figure*}[t]
\centering
\includegraphics[width=0.8\textwidth]{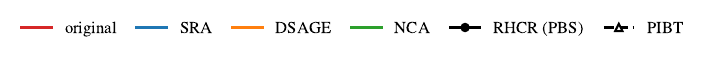}\\
\begin{subfigure}{0.24\textwidth}
    \includegraphics[width=1\textwidth]{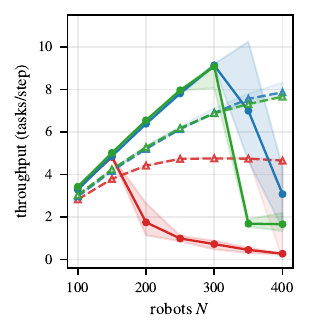}
    \caption{$w=1$}
\end{subfigure}%
\begin{subfigure}{0.24\textwidth}
    \includegraphics[width=1\textwidth]{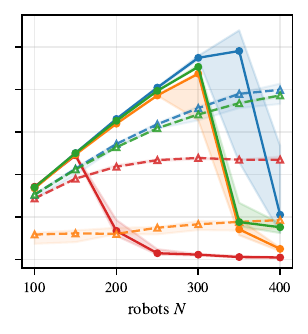}
    \caption{$w=5$}
\end{subfigure}%
\begin{subfigure}{0.24\textwidth}
    \includegraphics[width=1\textwidth]{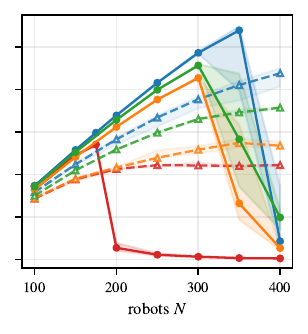}
    \caption{$w=10$}
\end{subfigure}%
\begin{subfigure}{0.24\textwidth}
    \includegraphics[width=1\textwidth]{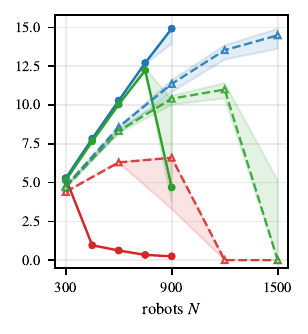}
    \caption{the $66\times69$ warehouse}
    \label{fig:scale2x}
\end{subfigure}
\caption{Throughput against robot count $N$: medians and interquartile ranges.
(a)--(c): the $33\times36$ warehouse per demand skew, over $10$ RHCR (PBS) and $5$ PIBT simulation seeds per layout, SRA pooling its annealing seeds.
(d): the $66\times69$ warehouse at $w=10$, over $3$ simulation seeds per layout, SRA pooling $3$ annealing seeds and NCA pooling $5$ generated layouts per planner; the RHCR (PBS) points at $N=300$ for the original and SRA layouts are single runs.}
\label{fig:sweep}
\end{figure*}
\paragraph{Search cost.}
\Cref{fig:convergence} overlays the best-so-far throughput of SRA and baselines in all three skews.
The baselines' searches cost $25{,}000$ simulations and as much as $25$ hours on a 64-core machine, while SRA runs on a single CPU core.
SRA replaces the per-candidate simulation with a stress-field evaluation that takes $0.3$ seconds: $3500$ evaluations, zero simulations, and $19$ minutes on one CPU core.
With RHCR, SRA surpasses the final quality of the baselines in $8$ minutes at $w=10$ and $3$ minutes at $w=5$.
The PIBT results show a similar trend, with SRA being significantly faster than the baselines while maintaining similar or better throughput.

\paragraph{The gain is planner-independent.}
DSAGE and NCA require separate optimizations with RHCR and PIBT, while SRA is planner-agnostic: we can optimize the layouts once and deploy different planners.
The PIBT block of Table~\ref{tab:external} confirms this at every skew: the original layout saturates at $\lambda=4.5$--$4.8$, and the SRA layouts reach $\lambda=7.2$--$7.8$.
Appendix~A.2 repeats the comparison under a third planner, namely RHCR with ECBS~\cite{barer2014suboptimal}.

\paragraph{SRA makes lifelong MAPF scale to more robots.}
\label{sec:exp-staircase}
\Cref{fig:sweep} compares throughput across various numbers of robots: the best layouts of the baselines follow the SRA layouts up to $N=250$ and fall behind at $N=300$--$350$ in every skew.
With $w{=}10$, the original layout climbs to a median $\lambda=5.4$ at $N=175$; at $N=200$ eight of ten runs end below $1$, and every run at $N\ge250$ ends below $0.4$.
The SRA layouts do not collapse at $N=300$ (median $\lambda=9.72$; one of the $50$ runs ends at $\lambda=2.9$) and hold until $N=350$, where $12$ of $50$ runs end between $1$ and $6$ and the survivors reach a median $\lambda\approx11$.
The same planner sustains roughly twice as many robots, from $N\approx175$ to $N\approx300$--$350$.

\paragraph{Visualization of optimized layouts.}
\label{sec:exp-gen}
\begin{figure}[t]
\centering
\includegraphics[width=0.9\columnwidth]{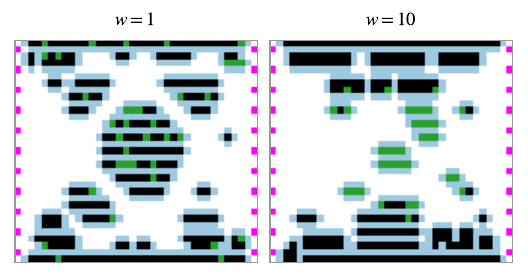}
\caption{The SRA layouts at $w=1$ and $w=10$ (annealing seed $0$).}
\label{fig:layouts}
\end{figure}
\Cref{fig:layouts} renders the SRA layouts at $w=1$ and $w=10$ (annealing seed $0$); Appendix~A.3 renders one per skew.
The structure adapts visibly: at $w=10$ the high-frequency shelves form short corridors in the center, and the low-frequency mass moves to the top and bottom edges.
The bottleneck load does not change with the skew: while the skew spans $10\times$, the optimized $\ell^*$ stays within $[0.058, 0.072]$.

\paragraph{Generalization across workstation demands.}
$w$ only controls demands at the shelves. We next vary the workstation demands $\omega_j$: the left-margin workstations are visited twice as often.
SRA stays unchanged: every flow of \Cref{eq:F} already carries the workstation weight $\omega_j$. The results are shown in Appendix~A.4.
The original layout collapses at $N=200$.
For the SRA layouts, $37$ of $50$ runs at $N=250$ (five annealing seeds, 10 simulation seeds each) end above $\lambda=8$ with median $\lambda=8.4$; at $N=300$ they collapse.
The high-frequency shelves end closer to the left margin: their average distance from it falls from $17.5$ columns in the original layout---the center of the $36$-column-wide grid---to $9.4$ columns in the annealed layouts.

\paragraph{Generalization to a larger warehouse.}
Neither baseline optimizes at the scale of the $66\times69$ warehouse (Sec.~\ref{sec:setup}): both searches run on the $33\times36$ map, and NCA then generates larger layouts with the trained generator, without further optimization~\cite{ZhangNCA2023}.
SRA optimizes the $66\times69$ warehouse directly: the anneal lowers $\ell^*$ from $0.084$ to $0.033$, in $12$ hours on one CPU core without simulation.
Figure~\ref{fig:scale2x} sweeps the robot count: under RHCR (PBS) the original layout collapses by $N=450$, while the SRA layouts reach $N=900$ with zero collapses in $36$ runs (median $\lambda=14.9$).
Under PIBT the original layout saturates at $\lambda\approx6$ and deadlocks from $N=1200$ on; the SRA layouts reach $\lambda=14.4$ at $N=1500$ and are still climbing.
The NCA-generated layouts follow the SRA layouts to $N=750$ under RHCR (PBS) and to $N=1200$ under PIBT; at $N=900$ and $N=1500$ they fall to median $\lambda=4.7$ and $\lambda=0$ (Appendix~C).

\subsection{Why SRA Works: The Stress Field Is Faithful}
\label{sec:exp-faithful}

Having shown that SRA outperforms the baselines, we now ask why.
At $w=10$, the right column of \Cref{fig:teaser} compares the predicted field with the traffic realized by an RHCR (PBS) run: the Spearman correlation between the predicted stress and the realized traversal counts is $\rho=0.91$ on the original layout and $0.81$ on the SRA optimized layout.
The corridors the field predicts to be congested are the ones that congest in simulation.
The bottleneck law sharpens this into a number.
Calibrating over RHCR simulations on this map---spanning layouts and robot counts---gives the single-vertex capacity $c=0.73$; the same value is used for all layouts.
Throughput then cannot exceed $c/\ell^*$, which gives $5.2$ for the original layout and $12$ for the SRA layouts.
PBS collapses instead of saturating: its throughput touches the ceiling at $N=175$ and drops to zero past it, never holding a plateau.
PIBT does not collapse: swept up to $N=400$ (\Cref{fig:sweep}, dashed), it stays flat from $N=250$ on at $\lambda\approx4.4$ on the original layout, $85\%$ of the ceiling calibrated on RHCR runs.
On the SRA layouts throughput is still climbing at $8.8$ when the sweep ends.
The saturation plateau is the bottleneck law made visible.
The collapse points of \Cref{fig:sweep} show the same law: SRA lowers $\ell^*$ from $0.139$ to $\approx0.060$, the collapse point moves with the ceiling, and the $N=350$ survivors sit just beneath $c/\ell^*$.

Finally, we ask how far the field can be trusted: we continue annealing well past the point where throughput stops improving and show the results in Appendix~A.5.
Simulating the intermediate layouts shows throughput saturating by roughly step $1500$, while $\ell^*$ falls another $19\%$ over the remaining $2000$ steps; collapse counts across the snapshots show no trend.
The field is a guide for the search, not a predictor of throughput, and early stopping is safe.

\subsection{Ablation Study}
\label{sec:exp-ablation}

All ablations run at $w=10$: five annealing seeds per variant, each evaluated under RHCR (PBS) at $N=300$ with $10$ simulation seeds, following the evaluation protocol of Sec.~\ref{sec:setup}.

\subsubsection{Necessity of Congestion Term} To demonstrate the necessity of the congestion term in \Cref{eq:Energy}, we remove it entirely.
Setting $\alpha=0$ reduces the energy to pure travel distance, similar to the prior analytical layout optimization methods mentioned in Sec.~\ref{sec:layoutopt}.
The resulting layouts pack the high-frequency shelves into one dense central band, where the mean distance to the workstations is smallest (rendered in Appendix~A.6), and deadlock without exception: all $50$ runs at $N=250$ and all $50$ at $N=300$ collapse, none above $\lambda=0.7$.
The full energy turns the same anneal into layouts that reach mean $\lambda=9.57$.
Distance optimization does not merely miss the congestion gains; it manufactures collapse.

\subsubsection{Target Selection} We then ablate the relocation rules, under an equal step budget, keeping the energy and the Metropolis rule fixed.
We compare (1) \emph{relocate} (ours), which places the source shelf at the best vertex map-wide, (2) \emph{hop}, which keeps the source rule but restricts the target to the eight vertices surrounding the source shelf's current position (the four edge-adjacent and the four diagonal ones), and (3) \emph{random}, which draws both the source shelf and the target vertex uniformly at random.
At $N=300$, relocate reaches mean $\lambda=9.57$ with $1$ collapse in $50$ runs; random reaches mean $\lambda=8.88$ with none; hop reaches only mean $\lambda=6.48$ and collapses in $20$ of $50$ runs.
Random directs nothing and still beats hop, so what the gain needs most is the map-wide reach of the relocation; all the guidance together adds the remaining $0.7$.
Hop gets trapped in configurations that keep the bottleneck.
\subsubsection{Source Selection}
Replacing the felt-stress rule of \texttt{getSource} by a uniformly random source changes neither $\ell^*$ nor throughput (Appendix~A.7): the felt score is kept for interpretability, not for quality.

\section{Conclusion and Future Work}
\label{sec:conclusion}

In this paper, we presented Stress-Relief Annealing (SRA), a polynomial-time simulation-free layout optimization algorithm for automated warehouses.
SRA turns the task demand into a per-vertex stress field whose peak $\ell^*$ caps steady-state throughput at $c/\ell^*$.
Guided by this field, SRA relocates shelves to lower $\ell^*$ while keeping trips short.
Our experiments show that SRA makes lifelong MAPF scale from $N\approx175$ on a human-designed warehouse to $N\approx300$--$350$ robots, matches or outperforms the evolutionary baselines in $19$ minutes on one CPU core against their $25$ hours on a $64$-core machine, and generalizes across MAPF planners, demand structures, and a much larger warehouse.

Our work is limited in several ways.
First, the single-vertex capacity $c$ is calibrated once per simulator setup, and the bound it feeds is validated qualitatively (lower $\ell^*$, later collapse), with no claim of tightness.
Second, our prototype recomputes the field per endpoint, at $O(|S|\,M\,|V|)$ per step instead of the $O(M\,|V|)$ of the Brandes-style accumulation (Sec.~\ref{sec:surrogate}), which we have not implemented; on this prototype the anneal takes $19$ minutes on the original warehouse and $12$ hours on the $66\times69$ one.

Future work is to scale to warehouses several times larger, building on that accumulation, on incremental field updates after a single relocation, and on batched relocations.

\bibliography{aaai2027}

\begin{thebibliography}{34}
\providecommand{\natexlab}[1]{#1}

\bibitem[{Barer et~al.(2014)Barer, Sharon, Stern, and Felner}]{barer2014suboptimal}
Barer, M.; Sharon, G.; Stern, R.; and Felner, A. 2014.
\newblock Suboptimal variants of the conflict-based search algorithm for the multi-agent pathfinding problem.
\newblock In \emph{Proceedings of the International Symposium on Combinatorial Search (SoCS)}, 19--27.

\bibitem[{Bhatt et~al.(2022)Bhatt, Tjanaka, Fontaine, and Nikolaidis}]{bhatt2022dsage}
Bhatt, V.; Tjanaka, B.; Fontaine, M.; and Nikolaidis, S. 2022.
\newblock Deep Surrogate Assisted Generation of Environments.
\newblock In \emph{Proceedings of the Advances in Neural Information Processing Systems (NeurIPS)}, 37762--37777.

\bibitem[{Brandes(2001)}]{brandes2001faster}
Brandes, U. 2001.
\newblock A faster algorithm for betweenness centrality.
\newblock \emph{Journal of Mathematical Sociology}, 25(2): 163--177.

\bibitem[{Brown(2022)}]{Brown2023amazonrobot}
Brown, A.~S. 2022.
\newblock How {Amazon} Robots Navigate Congestion.
\newblock \url{https://www.amazon.science/latest-news/how-amazon-robots-navigate-congestion}.
\newblock Accessed: 2023-05-09.

\bibitem[{Chen et~al.(2024)Chen, Harabor, Li, and Stuckey}]{chen2024traffic}
Chen, Z.; Harabor, D.; Li, J.; and Stuckey, P.~J. 2024.
\newblock Traffic flow optimisation for lifelong multi-agent path finding.
\newblock In \emph{Proceedings of the AAAI Conference on Artificial Intelligence (AAAI)}, 20674--20682.

\bibitem[{Ewing et~al.(2022)Ewing, Ren, Kansara, Sathiyanarayanan, and Ayanian}]{ewing2022betweenness}
Ewing, E.; Ren, J.; Kansara, D.; Sathiyanarayanan, V.; and Ayanian, N. 2022.
\newblock Betweenness Centrality in Multi-Agent Path Finding.
\newblock In \emph{Proceedings of the International Conference on Autonomous Agents and Multiagent Systems (AAMAS)}, 400--408.

\bibitem[{Fontaine and Nikolaidis(2023)}]{fontaine2023cmamae}
Fontaine, M.; and Nikolaidis, S. 2023.
\newblock Covariance matrix adaptation map-annealing.
\newblock In \emph{Proceedings of the Genetic and Evolutionary Computation Conference (GECCO)}, 456--465.

\bibitem[{Freeman(1977)}]{freeman1977betweenness}
Freeman, L.~C. 1977.
\newblock A set of measures of centrality based on betweenness.
\newblock \emph{Sociometry}, 40(1): 35--41.

\bibitem[{Han and Yu(2022)}]{han2022space}
Han, S.~D.; and Yu, J. 2022.
\newblock Optimizing Space Utilization for More Effective Multi-Robot Path Planning.
\newblock In \emph{Proceedings of the International Conference on Robotics and Automation (ICRA)}, 10709--10715.

\bibitem[{Hönig et~al.(2019)Hönig, Kiesel, Tinka, Durham, and Ayanian}]{honig2019warehouse}
Hönig, W.; Kiesel, S.; Tinka, A.; Durham, J.~W.; and Ayanian, N. 2019.
\newblock Persistent and Robust Execution of {MAPF} Schedules in Warehouses.
\newblock \emph{IEEE Robotics and Automation Letters}, 4(2): 1125--1131.

\bibitem[{Kirkpatrick, Gelatt~Jr, and Vecchi(1983)}]{kirkpatrick1983optimization}
Kirkpatrick, S.; Gelatt~Jr, C.~D.; and Vecchi, M.~P. 1983.
\newblock Optimization by simulated annealing.
\newblock \emph{science}, 220(4598): 671--680.

\bibitem[{Kou et~al.(2020)Kou, Peng, Ma, Kumar, and Koenig}]{KouAAAI20}
Kou, N.~M.; Peng, C.; Ma, H.; Kumar, T. K.~S.; and Koenig, S. 2020.
\newblock Idle Time Optimization for Target Assignment and Path Finding in Sortation Centers.
\newblock In \emph{Proceedings of the {AAAI} Conference on Artificial Intelligence (AAAI)}, 9925--9932.

\bibitem[{Lamballais, Roy, and De~Koster(2017)}]{lamballais2017estimating}
Lamballais, T.; Roy, D.; and De~Koster, M. 2017.
\newblock Estimating performance in a robotic mobile fulfillment system.
\newblock \emph{European Journal of Operational Research}, 256(3): 976--990.

\bibitem[{Lazowska et~al.(1984)Lazowska, Zahorjan, Graham, and Sevcik}]{lazowska1984quantitative}
Lazowska, E.~D.; Zahorjan, J.; Graham, G.~S.; and Sevcik, K.~C. 1984.
\newblock \emph{Quantitative system performance: computer system analysis using queueing network models}.
\newblock Prentice-Hall, Inc.

\bibitem[{Li et~al.(2021)Li, Tinka, Kiesel, Durham, Kumar, and Koenig}]{li2021lifelong}
Li, J.; Tinka, A.; Kiesel, S.; Durham, J.~W.; Kumar, T.~S.; and Koenig, S. 2021.
\newblock {Lifelong multi-agent path finding in large-scale warehouses}.
\newblock In \emph{Proceedings of the AAAI Conference on Artificial Intelligence (AAAI)}, 11272--11281.

\bibitem[{Liu et~al.(2019)Liu, Ma, Li, and Koenig}]{LiuAAMAS19}
Liu, M.; Ma, H.; Li, J.; and Koenig, S. 2019.
\newblock Task and Path Planning for Multi-Agent Pickup and Delivery.
\newblock In \emph{Proceedings of the International Conference on Autonomous Agents and Multi-Agent Systems (AAMAS)}, 1152--1160.

\bibitem[{Ma et~al.(2019)Ma, Harabor, Stuckey, Li, and Koenig}]{ma2019searching}
Ma, H.; Harabor, D.; Stuckey, P.~J.; Li, J.; and Koenig, S. 2019.
\newblock Searching with consistent prioritization for multi-agent path finding.
\newblock In \emph{Proceedings of the AAAI Conference on Artificial Intelligence (AAAI)}, 7643--7650.

\bibitem[{Ma et~al.(2017)Ma, Li, Kumar, and Koenig}]{ma2017lifelong}
Ma, H.; Li, J.; Kumar, T.~S.; and Koenig, S. 2017.
\newblock Lifelong Multi-Agent Path Finding for Online Pickup and Delivery Tasks.
\newblock In \emph{Proceedings of the Conference on Autonomous Agents and MultiAgent Systems (AAMAS)}, 837--845.

\bibitem[{Metropolis et~al.(1953)Metropolis, Rosenbluth, Rosenbluth, Teller, and Teller}]{metropolis1953equation}
Metropolis, N.; Rosenbluth, A.~W.; Rosenbluth, M.~N.; Teller, A.~H.; and Teller, E. 1953.
\newblock Equation of state calculations by fast computing machines.
\newblock \emph{The journal of chemical physics}, 21(6): 1087--1092.

\bibitem[{Mouret and Clune(2015)}]{mouret2015illuminating}
Mouret, J.-B.; and Clune, J. 2015.
\newblock Illuminating search spaces by mapping elites.
\newblock \emph{arXiv preprint arXiv:1504.04909}.

\bibitem[{Okumura et~al.(2022)Okumura, Machida, D{\'e}fago, and Tamura}]{okumura2022priority}
Okumura, K.; Machida, M.; D{\'e}fago, X.; and Tamura, Y. 2022.
\newblock Priority inheritance with backtracking for iterative multi-agent path finding.
\newblock \emph{Artificial Intelligence}, 310: 103752.

\bibitem[{Sheffi(1985)}]{sheffi1985urban}
Sheffi, Y. 1985.
\newblock \emph{Urban transportation networks}, volume~6.
\newblock Prentice-Hall, Englewood Cliffs, NJ.

\bibitem[{Stern et~al.(2019)Stern, Sturtevant, Felner, Koenig, Ma, Walker, Li, Atzmon, Cohen, Kumar, Boyarski, and Bart{\'{a}}k}]{Stern2019benchmark}
Stern, R.; Sturtevant, N.~R.; Felner, A.; Koenig, S.; Ma, H.; Walker, T.~T.; Li, J.; Atzmon, D.; Cohen, L.; Kumar, T. K.~S.; Boyarski, E.; and Bart{\'{a}}k, R. 2019.
\newblock Multi-Agent Pathfinding: Definitions, Variants, and Benchmarks.
\newblock In \emph{Proceedings of the International Symposium on Combinatorial Search (SoCS)}, 151--159.

\bibitem[{Wan et~al.(2018)Wan, Gu, Sun, Chen, Huang, and Jia}]{WanICARCV18}
Wan, Q.; Gu, C.; Sun, S.; Chen, M.; Huang, H.; and Jia, X. 2018.
\newblock Lifelong Multi-Agent Path Finding in a Dynamic Environment.
\newblock In \emph{Proceedings of the International Conference on Control, Automation, Robotics and Vision (ICARCV)}, 875--882.

\bibitem[{Wang and Botea(2011)}]{WangB11}
Wang, K.; and Botea, A. 2011.
\newblock {MAPP}: A Scalable Multi-Agent Path Planning Algorithm with Tractability and Completeness Guarantees.
\newblock \emph{Journal of Artificial Intelligence Research}, 42: 55--90.

\bibitem[{Wang and Botea(2008)}]{WangICAPS08}
Wang, K.~C.; and Botea, A. 2008.
\newblock Fast and Memory-Efficient Multi-Agent Pathfinding.
\newblock In \emph{Proceedings of the International Conference on Automated Planning and Scheduling (ICAPS)}, 380--387.

\bibitem[{Wardrop(1952)}]{wardrop1952road}
Wardrop, J.~G. 1952.
\newblock Road paper. some theoretical aspects of road traffic research.
\newblock \emph{Proceedings of the institution of civil engineers}, 1(3): 325--362.

\bibitem[{Wu et~al.(2020)Wu, Chi, Wang, and Wu}]{wu2020research}
Wu, S.; Chi, C.; Wang, W.; and Wu, Y. 2020.
\newblock Research of the layout optimization in robotic mobile fulfillment systems.
\newblock \emph{International Journal of Advanced Robotic Systems}, 17(6): 1729881420978543.

\bibitem[{Wurman, D'Andrea, and Mountz(2007)}]{wurman2008coordinating}
Wurman, P.~R.; D'Andrea, R.; and Mountz, M. 2007.
\newblock Coordinating Hundreds of Cooperative, Autonomous Vehicles in Warehouses.
\newblock In \emph{Proceedings of the {AAAI} Conference on Artificial Intelligence (AAAI)}, 1752--1760.

\bibitem[{Yang et~al.(2021)Yang, Liu, Feng, Zhang, and Qi}]{yang2021non}
Yang, X.; Liu, X.; Feng, L.; Zhang, J.; and Qi, M. 2021.
\newblock Non-traditional layout design for robotic mobile fulfillment system with multiple workstations.
\newblock \emph{Algorithms}, 14(7): 203.

\bibitem[{Yu and Wolf(2023)}]{Yu2023}
Yu, G.; and Wolf, M. 2023.
\newblock Congestion prediction for large fleets of mobile robots.
\newblock In \emph{Proceedings of the International Conference on Robotics and Automation (ICRA)}, 7642--7649.

\bibitem[{Zhang et~al.(2023{\natexlab{a}})Zhang, Fontaine, Bhatt, Nikolaidis, and Li}]{ZhangNCA2023}
Zhang, Y.; Fontaine, M.~C.; Bhatt, V.; Nikolaidis, S.; and Li, J. 2023{\natexlab{a}}.
\newblock Arbitrarily Scalable Environment Generators via Neural Cellular Automata.
\newblock In \emph{Proceedings of the Advances in Neural Information Processing Systems (NeurIPS)}, 57212--57225.

\bibitem[{Zhang et~al.(2023{\natexlab{b}})Zhang, Fontaine, Bhatt, Nikolaidis, and Li}]{zhangLayout23}
Zhang, Y.; Fontaine, M.~C.; Bhatt, V.; Nikolaidis, S.; and Li, J. 2023{\natexlab{b}}.
\newblock Multi-Robot Coordination and Layout Design for Automated Warehousing.
\newblock In \emph{Proceedings of the International Joint Conference on Artificial Intelligence (IJCAI)}, 5503--5511.

\bibitem[{Zhang et~al.(2022)Zhang, Fontaine, Hoover, and Nikolaidis}]{Zhang2021DeepSA}
Zhang, Y.; Fontaine, M.~C.; Hoover, A.~K.; and Nikolaidis, S. 2022.
\newblock Deep Surrogate Assisted MAP-Elites for Automated Hearthstone Deckbuilding.
\newblock In \emph{Proceedings of the Genetic and Evolutionary Computation Conference (GECCO)}, 158--167.

\end{thebibliography}

\appendix

\section{Additional Experimental Results}

\subsection{Where the Stress Field Fails}
\begin{figure*}[t!]
\centering
\includegraphics[width=0.85\textwidth]{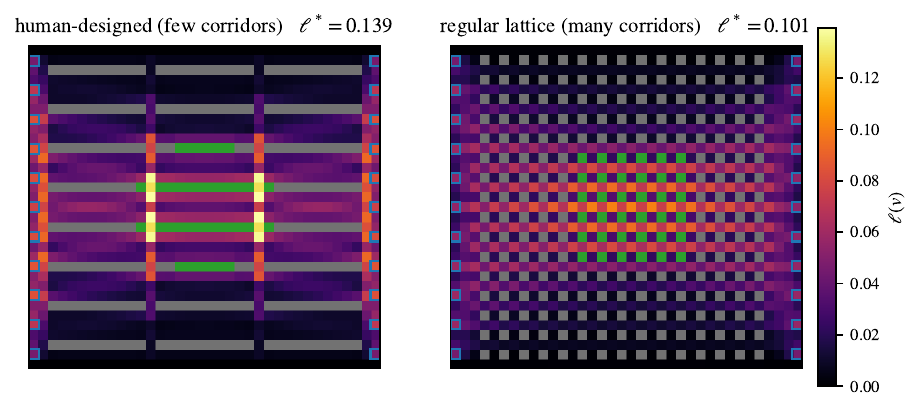}
\caption{Stress fields of the two start layouts: the human-designed baseline (left) funnels the flow into a few corridors, so its peak marks a binding bottleneck; the regular lattice (right) spreads the load over many equivalent corridors, so its lower peak no longer determines collapse.}
\label{fig:field-regimes}
\end{figure*}

Two layout families test where the field's guidance holds: layouts derived from the human-designed baseline, and layouts derived from a regular lattice.
On the human-designed family, the $\ell^*$ of an optimized layout correlates $+0.39$ (Spearman) with its count of simulations below $\lambda=7$: the field ranks fragility in the right direction.
On the lattice family the correlation inverts to $-0.25$.
The traffic prediction is not what fails: against the per-vertex traffic of the RHCR simulations that do not collapse, the field reads Spearman $0.80$ on the human-designed family and $0.87$ on the lattice family ($15$ layouts per family, $3$ simulation seeds).
What fails is the summary statistic: with many alternative corridors the load spreads over them (Figure~\ref{fig:field-regimes}), the most loaded vertex is no longer a binding bottleneck, and ranking layouts by the peak inverts.
Realistic warehouse baselines offer few alternative corridors, which is why the paper optimizes $\ell^*$ throughout; a statistic that predicts collapse when the load spreads remains open.

\subsection{A Third Planner}
\begin{table}[ht]
\small
\centering
\caption{Throughput $\lambda$ at $N=300$, $w=10$ under three planners: mean $\pm$ one standard deviation.
The original layout is evaluated over $10$ simulations; the SRA row pools the $5$ annealing seeds, $50$ simulations in all, and therefore sits below the best-layout values of Table~1 of the main paper.}
\label{tab:planners}
\begin{tabular}{lccc}
\toprule
 & RHCR (PBS) & RHCR (ECBS) & PIBT \\
\midrule
original & $0.13 \pm 0.06$ & $3.65 \pm 0.06$ & $4.45 \pm 0.14$ \\
SRA & $9.57 \pm 0.98$ & $8.06 \pm 0.17$ & $7.33 \pm 0.46$ \\
\bottomrule
\end{tabular}
\end{table}
Table~1 of the main paper evaluates the layouts under RHCR (PBS) and PIBT; here RHCR also runs with the bounded-suboptimal ECBS (suboptimality factor $2.0$) as its solver, at $w=10$ and $N=300$ with $10$ simulation seeds per layout.
Table~\ref{tab:planners} lists all three.
The SRA layouts stay above $\lambda=7$ under all three, achieving significantly better throughput than the original layout.
The anneal never ran ECBS, PBS, or PIBT; the same layouts serve all three.

\subsection{Demand Skew}
\begin{figure*}[t!]
\centering
\includegraphics[width=0.98\textwidth]{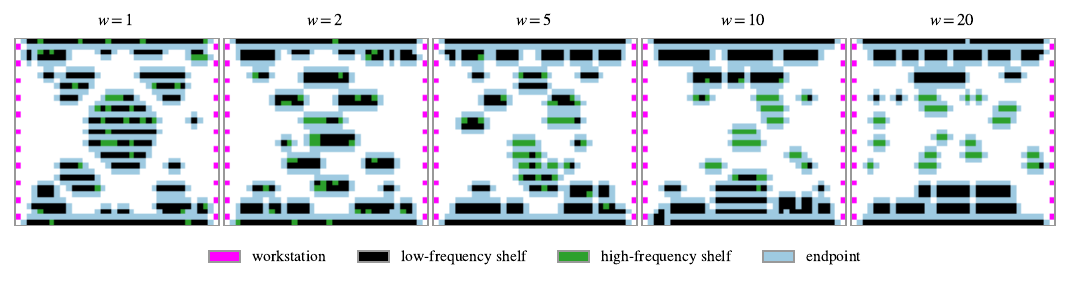}
\caption{The SRA layouts of Table~\ref{tab:skew} (annealing seed $0$), shown without the field overlay.}
\label{fig:skew-plain}
\end{figure*}

\begin{figure*}[t!]
\centering
\includegraphics[width=0.98\textwidth]{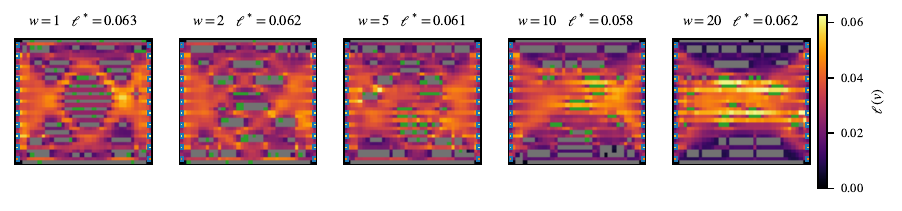}
\caption{The same layouts with their stress fields on a shared color scale.
Green vertices: high-frequency shelves; gray: low-frequency shelves; blue outlines: workstations.}
\label{fig:skew-fields}
\end{figure*}

Table~\ref{tab:skew} gives every layout and skew at $N=300$ ($3$ annealing seeds per skew, $10$ simulation seeds each).
The original layout collapses at every $w$; the SRA layouts collapse at none, and their lowest simulations at $w\le5$ congest at $\lambda=4.7$--$6.8$ without reaching zero.
Figure~\ref{fig:skew-plain} shows the five layouts; Figure~\ref{fig:skew-fields} overlays their stress fields.

\begin{table}[ht]
\centering
\caption{Throughput $\lambda$ at $N=300$ across demand skew, over $10$ simulations: mean $\pm$ one standard deviation.}
\label{tab:skew}
    \resizebox{1\linewidth}{!}{
\begin{tabular}{lcccc}
\toprule
 & $w{=}1$ & $w{=}2$ & $w{=}5$ & $w{=}20$ \\
\midrule
original & $0.73 \pm 0.32$ & $0.45 \pm 0.12$ & $0.24 \pm 0.10$ & $0.07 \pm 0.02$ \\
SRA $s_0$ & $8.79 \pm 0.94$ & $9.33 \pm 0.04$ & $9.16 \pm 0.82$ & $10.05 \pm 0.04$ \\
SRA $s_1$ & $8.07 \pm 1.70$ & $9.24 \pm 0.04$ & $9.50 \pm 0.08$ & $9.98 \pm 0.03$ \\
SRA $s_2$ & $9.25 \pm 0.05$ & $9.34 \pm 0.05$ & $9.58 \pm 0.06$ & $9.93 \pm 0.04$ \\
\bottomrule
\end{tabular}
}
\end{table}

\subsection{Non-Uniform Workstation Demands}
\begin{figure*}[t!]
\centering
\includegraphics[width=0.9\textwidth]{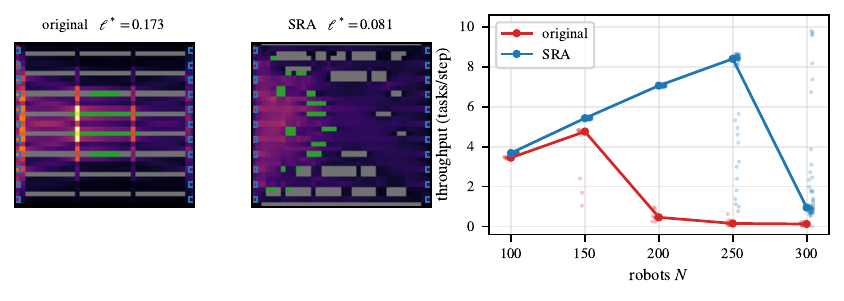}
\caption{Non-uniform workstation access at $w=10$: the left-margin workstations are visited twice as often.
Left and center: the weighted stress fields of the original layout and an SRA layout; the high-frequency shelves migrate toward the busy side.
Right: throughput against robot count $N$, median lines and individual simulations; the original collapses at $N=200$, the SRA layouts at $N=300$.}
\label{fig:wsstaircase-supp}
\end{figure*}
With the left-margin workstations visited twice as often, the original layout already collapses at $N=200$ (mean $\lambda=0.53$; Figure~\ref{fig:wsstaircase-supp}).
The SRA layouts do not collapse at $N=200$ (mean $\lambda=7.1$) or $N=250$ ($37$ of $50$ simulations, median $\lambda=8.4$; $10$ of the $13$ failures come from one annealing seed) and collapse at $N=300$ ($6$ of $50$ simulations do not collapse).
The concentrated load also limits how far $\ell^*$ can fall: it bottoms out at $0.080$ against $0.058$ in the uniform setting, which is the bottleneck-law explanation of the earlier collapse point.
The high-frequency shelves end closer to the left margin: their average distance from it falls from $17.5$ columns in the start layout---the center of the $36$-column-wide grid---to $9.4$ columns in the annealed layouts (mean over the $5$ anneals), without any instruction beyond the $\omega_j$ weights in the field.

\subsection{Annealing past Saturation}
\begin{figure}[t!]
\centering
\includegraphics[width=0.85\columnwidth]{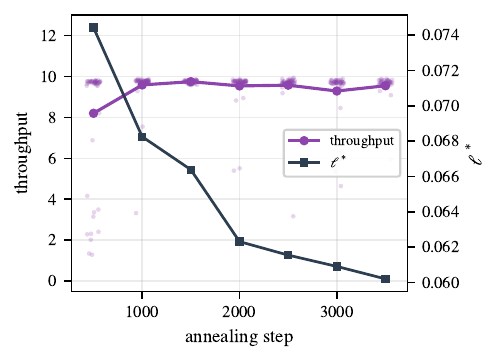}
\caption{Intermediate layouts along the anneal, simulated at $N=300$, $w=10$: individual simulations (dots), mean (line), and $\ell^*$ (right axis).
Throughput saturates by step $1500$; $\ell^*$ keeps falling.}
\label{fig:overopt-supp}
\end{figure}

Figure~\ref{fig:overopt-supp} renders the experiment of the main paper that continues annealing after throughput stops improving: past step $1500$, $\ell^*$ keeps falling while the simulated throughput stays flat.

\subsection{The Distance-Only Layouts}
\begin{figure*}[t!]
\centering
\includegraphics[width=0.98\textwidth]{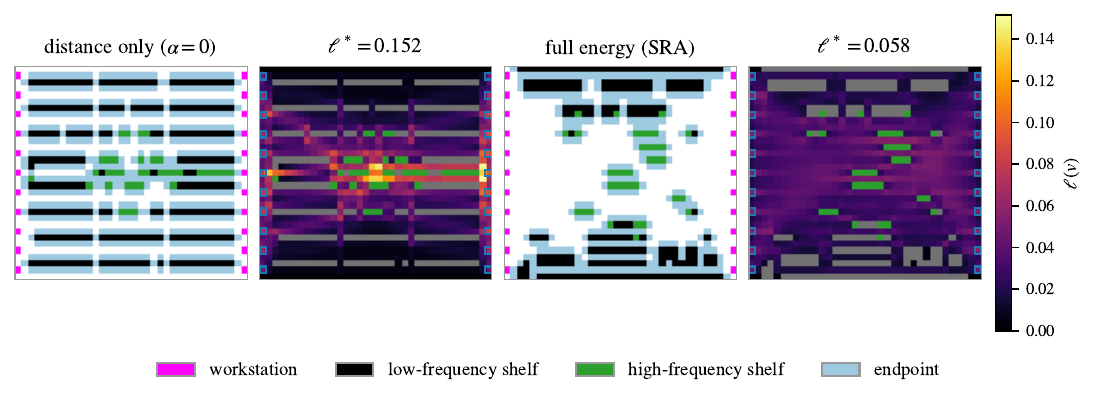}
\caption{A distance-only ($\alpha=0$) layout against a full-energy SRA layout ($w=10$, annealing seed $0$): each layout is followed by its stress field, on a shared color scale.}
\label{fig:distonly}
\end{figure*}

Figure~\ref{fig:distonly} renders a layout annealed with $\alpha=0$ against a layout annealed with the full energy ($w=10$, annealing seed $0$).
The distance term alone packs the high-frequency shelves into one dense central band, and the stress concentrates in the corridors that feed it: $\ell^*=0.152$ against $0.058$.
All $100$ simulations of the distance-only layouts collapse, none above $\lambda=0.7$.

\subsection{Source Selection}
We replace the softmax of Algorithm~\ref{alg:getsource} with a uniform draw over all shelves, keeping the placement rule and the energy unchanged.
The random-source variant matches the full method on both the surrogate and the simulator ($5$ annealing seeds, $3$ simulation seeds each).
Its final $\ell^*$ is $0.0583\pm0.0006$ against the full method's $0.0586\pm0.0004$ (Mann--Whitney $p=0.65$).
On the simulator, $4$ of its $15$ simulations finish below $\lambda=7$, against $5$ of $15$ for the full method.
What matters is the placement ranking; the felt score is a cheap tie-breaker kept for interpretability.

\section{Pseudocode of the Relocation Step}

Algorithms~\ref{alg:field}--\ref{alg:inittemp} give the four subroutines called in Algorithm~1 of the main paper.
Notation follows Sec.~4 of the main paper: $\rho$ is the expected occupancy, $D$ the mean distance to the workstations, $\alpha$ the weight of the congestion term, $w_s$ the demand weight of shelf $s$, $\partial s$ its endpoint set, $\omega_j$ the workstation weight, $d$ the shortest-path distance, $\sigma$ the shortest-path counts, $W_{\text{total}}$ the total demand weight, and $E$ the energy.
\begin{algorithm}[ht]
\caption{\texttt{computeStressField}$(G)$ --- compute $\ell^*$ and $L_w$}
\label{alg:field}
\begin{algorithmic}[1]
\FOR{each workstation $j$}
\STATE one BFS from $j$: the distance $d(v,j)$ and the shortest-path count $\sigma_{vj}$ for every vertex $v$
\ENDFOR
\STATE $L_w \gets$ the weighted trip length of Sec.~4 of the main paper, from the distances of line 2
\STATE $F(v) \gets 0$ for every vertex $v$
\FOR{each endpoint $e$, carrying $w_e \gets \sum_{s:\,e\in\partial s} w_s/|\partial s|$}
 \STATE one BFS from $e$: the distance $d(e,v)$ and the count $\sigma_{ev}$ for every vertex $v$
  \FOR{each workstation $j$ and each vertex $v$ with $d(e,v)+d(v,j)=d(e,j)$}
    \STATE $F(v) \gets F(v) + w_e\,\omega_j\,\sigma_{ev}\,\sigma_{vj}/\sigma_{ej}$
  \ENDFOR
\ENDFOR
\RETURN $\ell^* \gets \max_v F(v)/W_{\text{total}}$,\quad $L_w$
\end{algorithmic}
\end{algorithm}

Lines 6--11 are the per-endpoint recomputation named in the runtime paragraph of the main paper: one BFS and one $M\,|V|$ accumulation per endpoint, $O(|S|\,M\,|V|)$ in all.
\begin{algorithm}[ht]
\caption{\texttt{getSource}$(G)$ --- sample the shelf to relocate}
\label{alg:getsource}
\begin{algorithmic}[1]
\FOR{each shelf $s$ in $G$}
  \STATE $\mathrm{felt}(s) \gets \alpha\,\max_{e\in\partial s}\rho(e) + w_s\,\frac{1}{|\partial s|}\sum_{e\in\partial s} D(e)$
\ENDFOR
\STATE $z(s) \gets$ z-score of $\mathrm{felt}(s)$ over all shelves
\RETURN one shelf sampled with probability $\propto e^{z(s)}$
\end{algorithmic}
\end{algorithm}

\begin{algorithm}[ht]
\caption{\texttt{getTarget}$(G, s)$ --- first valid destination}
\label{alg:gettarget}
\begin{algorithmic}[1]
\FOR{each empty storage vertex $v$, in ascending $\mathrm{cost}(v)=\alpha\,\rho(v)+w_s\,D(v)$}
  \STATE $G' \gets G$ with $s$ relocated to $v$
  \IF{$G'$ is valid (Definition~2 of the main paper)}
    \RETURN $v$
  \ENDIF
\ENDFOR
\RETURN failure
\end{algorithmic}
\end{algorithm}
\begin{algorithm}[ht!]
\caption{\texttt{initializeTemperature}$(G)$ --- calibrate $T_0$}
\label{alg:inittemp}
\begin{algorithmic}[1]
\STATE $E_0 \gets E(G)$
\FOR{$i = 1$ \TO $40$}
  \IF{$10$ values recorded}
    \STATE \textbf{break}
  \ENDIF
  \STATE $s \gets \texttt{getSource}(G)$
  \STATE $v \gets \texttt{getTarget}(G, s)$ with the candidates restricted to the $8$ vertices surrounding $s$
  \IF{no candidate passes}
    \STATE \textbf{continue}
  \ENDIF
  \STATE $G' \gets G$ with $s$ relocated to $v$;\quad record $|E(G') - E_0|$ \COMMENT{$G'$ is discarded}
\ENDFOR
\RETURN $T_0 \gets$ the mean of the recorded values
\end{algorithmic}
\end{algorithm}
If no vertex passes the validity check, the step proposes no relocation.
The failure is specific to the sampled source shelf: the softmax of \texttt{getSource} can draw a different shelf at the next step, for which valid targets can exist.

Every trial in \texttt{initializeTemperature} relocates to one of the $8$ vertices surrounding the source shelf, whatever target rule the anneal itself uses, so the temperature scale does not depend on that choice.
\section{The $66\times69$ Warehouse}

\begin{figure*}[t!]
\centering
\includegraphics[width=0.85\textwidth]{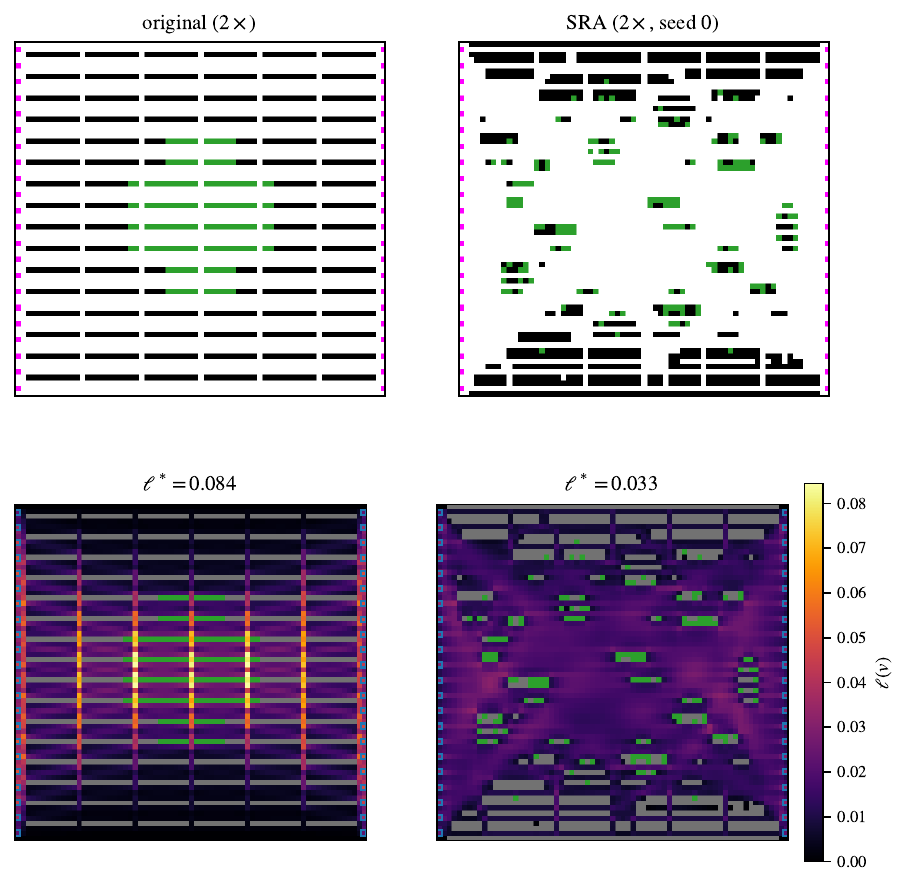}
\caption{The start layout of the $66\times69$ warehouse and the seed-$0$ SRA layout: layouts (top) and stress fields on a shared color scale (bottom).}
\label{fig:scale2x-renders}
\end{figure*}

This map doubles both dimensions of the warehouse of the main paper while preserving its structure: shelf rows every fourth row, $10$-wide shelf blocks separated by one-wide vertical aisles (six blocks per row), and a workstation every third row on both margins.
This gives a $66\times69$ grid with four times the shelves ($960$, density $0.211$ against $0.202$) and twice the workstations ($M=44$).
The high-frequency region is the center cross of the original map upsampled $2\times$: $144$ shelves, the same $15\%$ share.
The anneal runs the protocol of the main paper re-anchored on this map ($\alpha'=L_{w,0}/\ell^*_0$, $K=1$) at $N=1200$ for $10{,}000$ steps; it saturates near step $6{,}000$, consistent with the $4\times$ shelf count over the saturation near step $1{,}500$ on the original map.
Three annealing seeds give $\ell^*\in[0.032, 0.034]$.
Figure~\ref{fig:scale2x-renders} shows the start layout and one optimized layout.

\begin{figure*}[t!]
\centering
\includegraphics[width=0.98\textwidth]{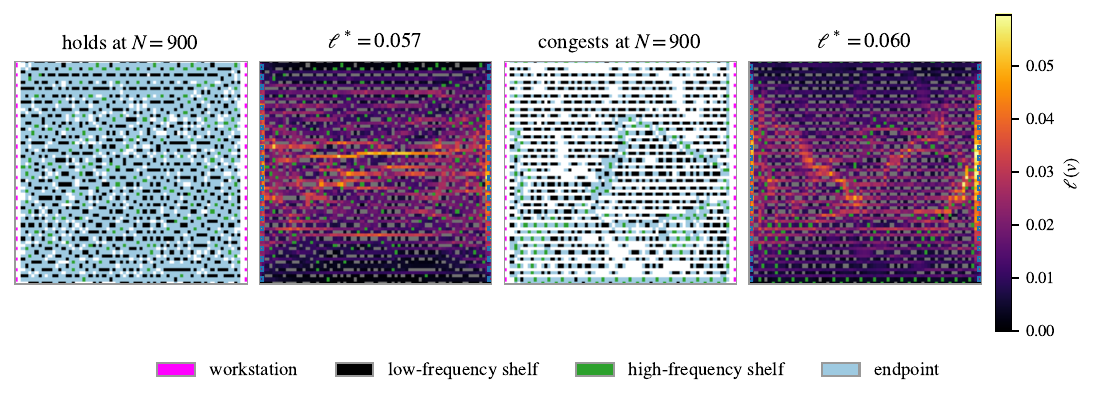}
\caption{Two NCA-generated $66\times69$ layouts: the one whose RHCR (PBS) simulations hold at $N=900$ and one whose simulations congest, each followed by its stress field, on a shared color scale.
Endpoints are the generators' own endpoint cells.}
\label{fig:nca2x-renders}
\end{figure*}

\paragraph{The NCA Generators at This Scale}
Each generator trained on the $33\times36$ warehouse grows a $66\times69$ layout in a single generator run, repaired by the same MILP as at training scale.
The RHCR (PBS) and PIBT sweeps run on these generated $66\times69$ layouts, $5$ per planner, from the $w=10$ searches behind Table~1 of the main paper: the RHCR-evaluated search for the RHCR (PBS) sweep, the PIBT-evaluated search for the PIBT sweep.
At $N=750$ under RHCR (PBS), $12$ of the $15$ simulations end at $\lambda=12.1$--$12.6$, against median $\lambda=12.7$ for SRA; at $N=900$, $11$ of the $15$ congest at $\lambda=2.9$--$6.5$ without reaching zero, and one generator stays at $\lambda=14.2$--$14.4$, the SRA level.
Under PIBT the generated layouts reach median $\lambda=11.0$ at $N=1200$; at $N=1500$, $11$ of the $15$ simulations exceed the wall clock, as the simulations of the original layout do from $N=1200$ on.
Figure~\ref{fig:nca2x-renders} renders the generated layout that holds at $N=900$ and one that congests, with their stress fields.
The $\ell^*$ of the $5$ generated layouts spans $[0.051, 0.060]$ against $[0.032, 0.034]$ for the annealing seeds, which is the bottleneck-law explanation of the earlier fall.

\section{Baseline Parameters}

DSAGE and NCA run the implementations released by their authors, configured on the warehouse of the main paper.
Both maximize throughput minus a Hamming-distance penalty (weight $5$) toward the human-designed start layout, evaluating each candidate with five RHCR (PBS) simulations of $1{,}000$ timesteps at $N=300$ under the target skew, with the same replanning period ($5$) and planning window ($10$) as the evaluations in the main paper.
Both are quality-diversity searches over a $100\times100$ archive spanning the number of shelf components and the layout entropy, with a budget of $5{,}000$ simulated candidates.
DSAGE proposes layouts by tile mutation (batch $50$) and accelerates the search with its convolutional surrogate, retrained on the accumulated simulations and exploited for $10{,}000$ archive iterations between simulation rounds.
NCA searches the $2{,}372$ weights of its generator network with CMA-MAE ($5$ CMA-ES emitters, batch $10$, $\sigma_0=0.2$), each layout grown from a fixed seed pattern in $50$ generator steps.
Per-run search times: DSAGE $19.2 \pm 1.0$ hours over $10$ runs on $64$ cores, NCA $29.6 \pm 1.1$ hours over $15$ runs, with NCA's four $16$-core runs rescaled to the $64$-core reference by the core ratio.
Six further DSAGE runs on $16$ cores took $33$--$37$ hours; rescaling them by the core ratio would place them below every measured $64$-core run, so we exclude them from the mean.
The PIBT rows of Table~1 of the main paper come from the same searches with the five evaluation simulations run by PIBT instead of RHCR (PBS): DSAGE $2.5 \pm 0.3$ hours and NCA $2.6 \pm 0.1$ hours, $15$ runs each.
\section{Compute Configuration}

SRA and every simulation reported in this paper---RHCR (PBS and ECBS) and PIBT, across all tables and figures---run on one desktop machine with an 8-core AMD Ryzen 7 9700X and 32 GB of RAM.
Each anneal uses a single core; the search times of $18.8$ minutes on the original warehouse and $12$ hours on the $66\times69$ warehouse are single-core times on this machine.
Evaluation simulations run as independent single-core processes, up to eight in parallel.

The DSAGE and NCA require massive parallelization, and therefore run on the following machines: (1) a local machine with a 64-core AMD Ryzen Threadripper 7980X CPU, 256GB of RAM, and an Nvidia GTX 1080Ti GPU, (2) a local machine with a 64-core AMD Ryzen Threadripper 3990X CPU and 128 GB of RAM, and (3) a high-performing cluster with numerous 64-core AMD EPYC 7742 CPUs, each with 256 GB of RAM.
\end{document}